\documentclass[useAMS,usenatbib]{mnras}

\usepackage{natbib}
\usepackage{mathtools}
\usepackage{amsmath}
\usepackage{amssymb}

\title[Globular cluster ages]{The SLUGGS Survey: Measuring globular cluster ages using both photometry and spectroscopy}

\author[Usher et al.]{Christopher~Usher,$^{1}$\thanks{email: c.g.usher@ljmu.ac.uk}
Jean~P.~Brodie,$^{2}$
Duncan~A.~Forbes,$^{3}$ \newauthor
Aaron~J.~Romanowsky,$^{4,2}$,
Jay~Strader,$^{5}$
Joel~Pfeffer,$^{1}$
Nate~Bastian$^{1}$\\
$^1$Astrophysics Research Institute, Liverpool John Moores University, 146 Brownlow Hill, Liverpool L3 5RF, UK \\
$^2$University of California Observatories, 1156 High Street, Santa Cruz, CA 95064, USA \\
$^3$Centre for Astrophysics \& Supercomputing, Swinburne University, Hawthorn VIC 3122, Australia \\
$^4$Department of Physics and Astronomy, San Jos\'e State University, One Washington Square, San Jose, CA 95192, USA \\
$^5$Department of Physics and Astronomy, Michigan State University, East Lansing, Michigan 48824, USA \\
}

\begin{document}
\maketitle

\begin{abstract}
Globular cluster ages provide both an important test of models of globular cluster formation and a powerful method to constrain the assembly history of galaxies.
Unfortunately, measuring the ages of unresolved old stellar populations has proven challenging.
Here, we present a novel technique that combines optical photometry with metallicity constraints from near-infrared spectroscopy in order to measure ages.
After testing the method on globular clusters in the Milky Way and its satellite galaxies, we apply our technique to three massive early-type galaxies using data from the SLUGGS Survey.
The three SLUGGS galaxies and the Milky Way show dramatically different globular cluster age and metallicity distributions, with NGC 1407 and the Milky Way showing mostly old globular clusters while NGC 3115 and NGC 3377 show a range of globular ages.
This diversity implies different galaxy formation histories and that the globular cluster optical colour--metallicity relation is not universal as is commonly assumed in globular cluster studies.
We find a correlation between the median age of the metal rich globular cluster populations and the age of the field star populations, in line with models where globular cluster formation is a natural outcome of high intensity star formation.
\end{abstract}


\section{Introduction}
 \label{sec:intro}
Globular clusters (GCs) have long held promise as probes of galaxy formation and evolution (see reviews by \citealt{2006ARA&A..44..193B} and \citealt{2018RSPSA.47470616F}).
Found in virtually all galaxies with stellar masses above $10^{9}$ M$_{\sun}$ and in most above $10^{8}$ M$_{\sun}$, the high surface brightnesses of GCs allow the old and intermediate age populations of galaxies to be studied at much greater distances than with individual stars.
Since most GCs in the Milky Way (MW) are old \citep[$\gtrsim 11$ Gyr, e.g.][]{2010ApJ...708..698D, 2013ApJ...775..134V}, GCs are commonly seen as fossils of the earlier stages of galaxy formation.
The properties of GCs observed today reflect both the conditions of GC formation but also the physics of GC survival since GCs lose mass and may even be destroyed by tidal interactions \citep[e.g.][]{2010ApJ...712L.184E, 2015MNRAS.454.1658K, 2019MNRAS.486.4030L}.
As such, by understanding how GCs form and evolve, we can extend our understanding of how individual galaxies form and evolve.

GC age distributions provide both a powerful test of GC formation models and a potentially important way to study galaxy assembly.
Two broad classes of models have been proposed for GC formation.
In the first class of models \citep[e.g.][]{1984ApJ...277..470P, 2014MNRAS.444.2377K, 2015ApJ...808L..35T, 2019ApJ...878L..23C} GC formation requires the special conditions of the early universe, forming before or during the epoch of reionization ($z \gtrsim 6$).
In the second class \citep[e.g.][]{1997ApJ...480..235E, 2010MNRAS.403L..36S, 2010ApJ...718.1266M, 2015MNRAS.454.1658K, 2017ApJ...834...69L, 2018MNRAS.480.2343C, 2018MNRAS.474.4232K, 2019arXiv190611261M}, GC formation is the natural outcome of intense star formation and GC formation should consequently peak just before the peak of cosmic star formation at redshift $z \sim 2$ \citep{2019MNRAS.482.4528E, 2019MNRAS.486.5838R}.
In the first class of models, GC ages should sharply peak around 13 Gyr and show relatively little variance galaxy-to-galaxy; in the second the age distribution should be broader and both the width and position of the peak should vary with galaxy assembly history \citep[e.g.][]{2019MNRAS.486.3134K}. 

Following a long history of using the GC system of the Milky Way (MW) to understand its formation \citep[e.g.][]{1978ApJ...225..357S, 2010MNRAS.404.1203F, 2013MNRAS.436..122L}, \citet{2019MNRAS.486.3180K} used the ages and metallicities of GCs in the MW together with predictions of the cosmological E-MOSAICS simulations \citep{2018MNRAS.475.4309P, 2019MNRAS.486.3134K} of galaxy and GC system formation to reconstruct in detail the assembly history of the MW.
\citet{2019MNRAS.486.3180K} placed strong constraints on the number, masses and redshifts of mergers that built up the MW and found that the MW assembled earlier than average for a galaxy of its mass.
While studies of the MW are important, it is only a single galaxy with a formation history that is likely atypical of a galaxy of its mass and environment \citep[e.g.][]{2018MNRAS.477.5072M}.
To test whether models for GC and galaxy formation reproduce the diverse population of galaxies we observe today we need to extend GC stellar population studies to a wider range of galaxies

Beyond the Local Group, GCs must be studied via their integrated light.
Unfortunately, the ages of old stellar populations are difficult to measure reliably from integrated light.
Optical colours suffer from a strong age--metallicity degeneracy \citep[e.g.][]{1994ApJS...95..107W}.
Adding near-infrared photometry can in principle break this degeneracy, although observing deep enough photometry over a wide area is challenging and has had limited success \citep[e.g.][]{2002A&A...391..453P, 2003A&A...405..487H, 2005A&A...443..413L, 2011A&A...525A..20C}.
Extragalactic GC ages have typically been measured using optical spectroscopy in the regions of the H$\beta$ and Mgb spectral features.
Unfortunately, the high signal-to-noise spectra required to disentangle the effects of age, metallicity and chemistry have limited these studies to relatively small samples of GCs.
These studies \citep[e.g.][]{1998ApJ...496..808C, 2001ApJ...563L.143F, 2005A&A...439..997P, 2005AJ....130.1315S, 2008MNRAS.385...40N} have generally found old ages ($\sim 12$ Gyr) but evidence of younger GCs has been seen in a few galaxies \citep[e.g.][]{2006ApJ...646L.107C, 2006MNRAS.372.1259S, 2008AJ....136..234C, 2008MNRAS.386.1443B, 2018MNRAS.473.2688M, 2018MNRAS.479..478S}.
A major systematic uncertainty for integrated light ages remains the horizontal branch (HB).
Stellar population models have trouble reproducing the ages of MW GCs with blue HBs \citep[e.g.][]{1994ApJS...95..107W, 2000ApJ...541..126M, 2007ApJS..171..146S, 2018ApJ...854..139C} as the presence of hot HB stars can mimic the appearance of a hotter (younger) main sequence turn-off.

In \citet{2015MNRAS.446..369U} we found clear evidence that the relationship between optical colour and the strength of the near-infrared calcium triplet (CaT) spectral feature varies between galaxies.
This suggests the stellar populations of GCs vary between galaxies (see also \citealt{2016ApJ...829L...5P} and \citealt{2019ApJ...879...45V}).
Recently \citet{2019MNRAS.482.1275U} studied the behaviour of the CaT in GCs in the Milky Way and its satellite galaxies and found that neither age or HB morphology has a measurable effect on the measured CaT strength for GCs older than a couple of Gyr.
By combining the age independent metallicity measurement from the CaT with optical photometry that is sensitive to both age and metallicity, we can break the degeneracy and measure both GC ages and metallicities.
In this paper we test this technique on GCs in the Milky Way and its satellite galaxies (Section \ref{sec:technique}) before applying it to three massive early-type galaxies from the SAGES Legacy Unifying Globulars and GalaxieS Survey \citep[SLUGGS;][]{2014ApJ...796...52B} (Section \ref{sec:sluggs}) and discussing our results in Section \ref{sec:discussion}.

\section{Ages from combining photometry and spectroscopy} \label{sec:technique}
To measure the ages of our globular clusters, we fit our photometry with stellar population synthesis models while using our CaT based [Fe/H] measurements as a constraint.
We used stellar population model grids calculated with version 3.0 of Flexible Stellar Population Synthesis (\textsc{FSPS}) code \citep{2009ApJ...699..486C, 2010ApJ...712..833C} using the MIST stellar isochrones \citep{2016ApJ...823..102C}, the MILES spectral library \citep{2006MNRAS.371..703S} and a \citet{2001MNRAS.322..231K} initial mass function.
These models were calculated at [Fe/H] = $[-3.0, -2.5, -2.0, -1.75, -1.50, -1.25, -1.00, -0.75, -0.50,\\ -0.25, +0.0, +0.5]$ and at ages between 0.1 and 15.8 Gyr in intervals of 0.05 dex.

Similarly to \citet{2017ApJ...837..170L} we performed a two step process to find the posterior distribution of our model for each GC.
First, we performed a coarse grid search in age and metallicity, calculating the best fitting mass for each age and metallicity given the input $E(B - V)$ reddening.
Second, we used version 2.2.1 of the Markov chain Monte Carlo (MCMC) code \textsc{emcee} \citep{2013PASP..125..306F} with 1000 walkers, 100 burn-in steps and 500 production steps to sample the posterior distribution.
We sampled the posteriors of age (assuming a flat prior between 0.1 and 15.8 Gyr), metallicity (assuming a flat prior between [Fe/H] $= -3.0$ and $+0.7$), stellar mass (assuming a flat prior in log mass) and reddening (assuming a Gaussian prior) subject to the constraints provided by each of the photometry bands and the CaT [Fe/H] measurements which are each assumed to be normally distributed.
We initialised the walkers in a small ball around the best fitting age, metallicity and mass from the grid search.
Due to the multi-modal nature of the likelihood function, we utilised the \textsc{emcee} parallel-tempered ensemble sampler rather than the default affine-invariant ensemble sampler.
When sampling the likelihood we linearly interpolated between the model grid points.
To correct for the effects of extinction we used the reddenings provided by \citet{2011ApJ...737..103S} for a $R_{V} = 3.1$ reddening law.
Our assumption of simple stellar populations is appropriate for GCs but may not be valid for stripped galactic nuclei (e.g. $\omega$ Cen, M54) which often show large metallicity spreads and extended star formation histories \citep[e.g.][]{2000AJ....119.1760L, 2010ApJ...722.1373J, 2010A&A...520A..95C, 2014ApJ...791..107V}. 

In this work we have used a relatively simple metallicity prior based on an empirical relationship between CaT strength and metallicity when fitting the photometry.
Our technique could be easily extended to use a more sophisticated age--metallicity (and possibly chemistry) prior from full spectral fitting \citep[e.g.][]{2008MNRAS.385.1998K, 2017A&A...601A..96L, 2017MNRAS.472.4297W, 2018ApJ...854..139C} or even by simultaneously fitting both photometry and spectroscopy to fully utilise the information contained in the spectroscopy \citep{2017ApJ...837..170L}.
A previous example of combining the stellar population constraints from optical photometry and spectroscopy can be found in \citet{2018ApJ...859...37G} who used both to place stronger constraints on the old ages of three ultra-diffuse galaxies.

\subsection{Tests on the Local Group Globular Clusters} \label{sec:local}

We test our method on observations of GCs in the MW and its satellite galaxies.
We use [Fe/H] measurements calculated from the CaT measurements given in table 5 of \citet{2019MNRAS.482.1275U} and equation 1 of \citet{2019MNRAS.482.1275U}. We note that the \citet{2019MNRAS.482.1275U} measurements were made using the same code and similar resolution spectra to our SLUGGS DEIMOS spectra.
For the MW GCs we use the $BVRI$ photometry from the 2010 edition of the \citet{1996AJ....112.1487H, 2010arXiv1012.3224H} catalogue, for the GCs in the Large and Small Magellanic Clouds we use the $BV$ photometry from \citet{1981A&AS...46...79V} and for the Fornax dwarf spheroidal GCs we use the $BV$ photometry from \citet{1969ApJS...19..145V}.
We note that $(B - V)$ and $(g - r)$ show similar behaviour with age and metallicity while the colours $(V - R)$, $(R - I)$, $(r - i)$ and $(i - z)$ all show similar behaviour to one another. 
Following \citet{1981A&AS...46...79V} we estimate an uncertainty of 0.02 mag in each of the photometric bands.
As in \citet{2019MNRAS.482.1275U}, we remove from our sample GCs whose CaT observations are based on observations with apertures enclosing less than $5 \times 10^{3}$ M$_{\sun}$ in order to reduce stochastic effects.
We use the reddening values from \citet{2019MNRAS.482.1275U} and assume the reddening uncertainty to be 0.02 mag added in quadrature with one tenth of the reddening.

\begin{figure}
    \includegraphics[width=240pt]{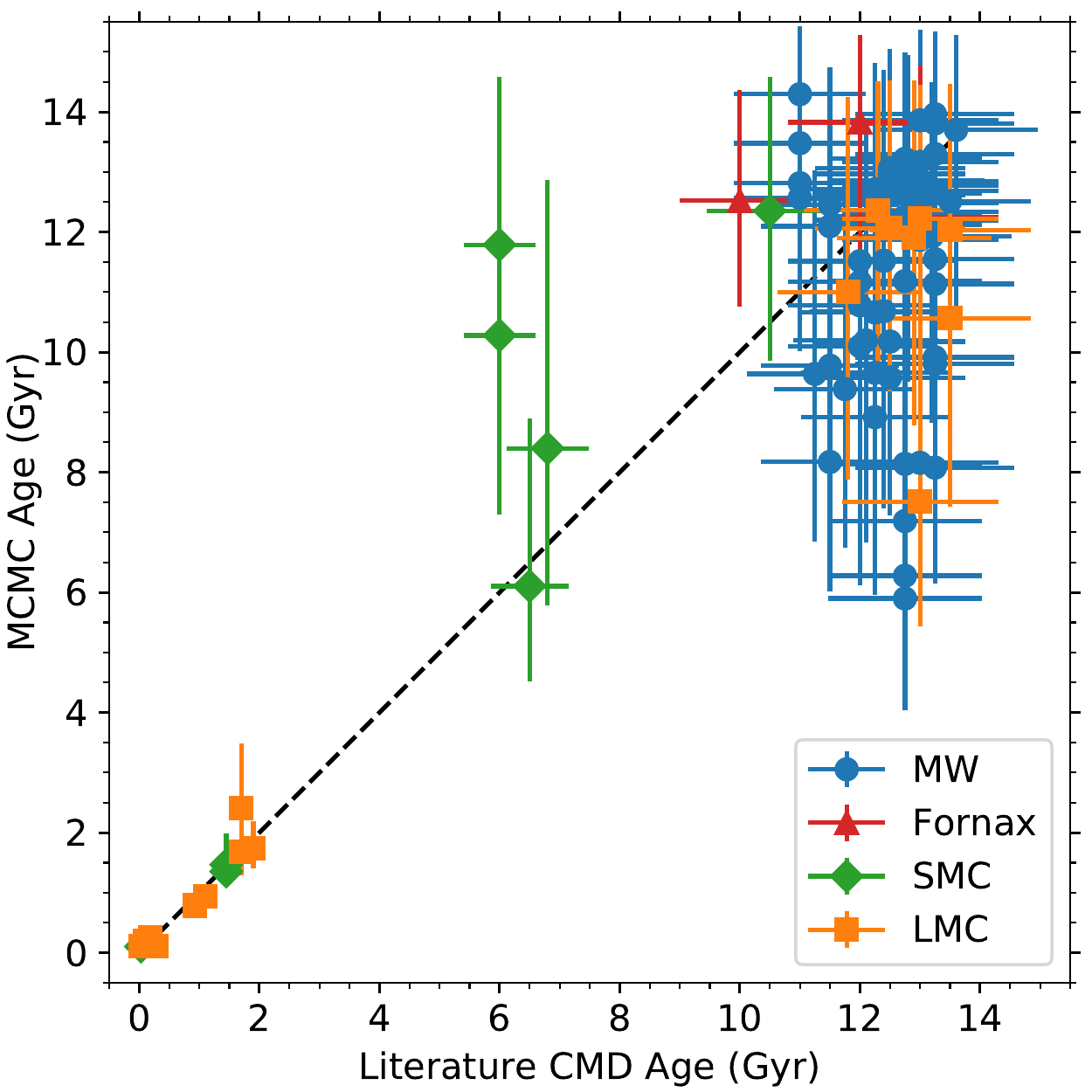}
    \caption{Medians of the MCMC age posteriors versus literature ages for GCs in the MW (blue circles), the LMC (orange squares), the SMC (green diamonds) and the Fornax dSph (red triangles).
    The dashed line is the one-to-one line.
    There is relatively good agreement for GCs of all ages both in the MW and in its satellite galaxies. \label{fig:age_comparison}}
\end{figure}

We plot the posterior distributions of one of our MW GCs (NGC~5272) in Figure \ref{fig:NGC5272_corner} and a comparison of the medians of our age posteriors to literature ages (from resolved colour magnitude diagrams) in Figure \ref{fig:age_comparison}. 
The sources of these ages are given in table 1 of \citet{2019MNRAS.482.1275U} and we refer the interested reader to section 2 of \citet{2017MNRAS.468.3828U} for further details.
For the MW GCs, these ages are mostly from \citet{2010ApJ...708..698D, 2011ApJ...738...74D} or were derived in a similar way \citep{2014ApJ...785...21M}.
The satellite galaxy GC ages are from a range of sources including \citet{1997AJ....114.1920G}, \citet{1998MNRAS.300..665O}, \citet{2008AJ....136.1703G}, \citet{2014ApJ...797...35G}, \citet{2015A&A...575A..62N} and \citet{2016A&A...590A..35D}.
In general, we see good agreement between our MCMC age constraints and literature ages.
Due to the upper limit on the age prior of 15.8 Gyr, the age posteriors of old GCs can extend to significantly younger ages than older ages.
At ages of $\sim 2$ Gyr and younger, where the CaT is a less reliable metallicity indicator \citep{2019MNRAS.482.1275U}, our technique still returns ages reliable at the 500 Myr level.
We note that at these ages the colours are relatively less sensitive to metallicity.
When the CaT metallicities show the largest discrepancies with the literature, the CaT metallicities are over estimated and result in younger ages.
The difference in our ages and literature ages shows no relationship with GC age, metallicity, mass or reddening.
Nor is there any significant correlation between the age difference and the HB parameter $\Delta(V - I$ of \citet{2010ApJ...708..698D} (a Kendell's $\tau$ of $-0.12$ for a $p$-value of 0.32) or the HB parameters $L1$ and $L2$ of \citet{2014ApJ...785...21M} ($\tau = -0.06$ and $-0.15$ respectively for $p = 0.64$ and $0.21$).
We note that the main limiting factor of these comparisons is the quality of the photometry and of the reddening correction as the uncertainties on the CaT metallicities are significantly smaller. 

\section{Applying the Technique to SLUGGS Galaxy Globular Clusters} \label{sec:sluggs}
\begin{figure*}
    \includegraphics[width=504pt]{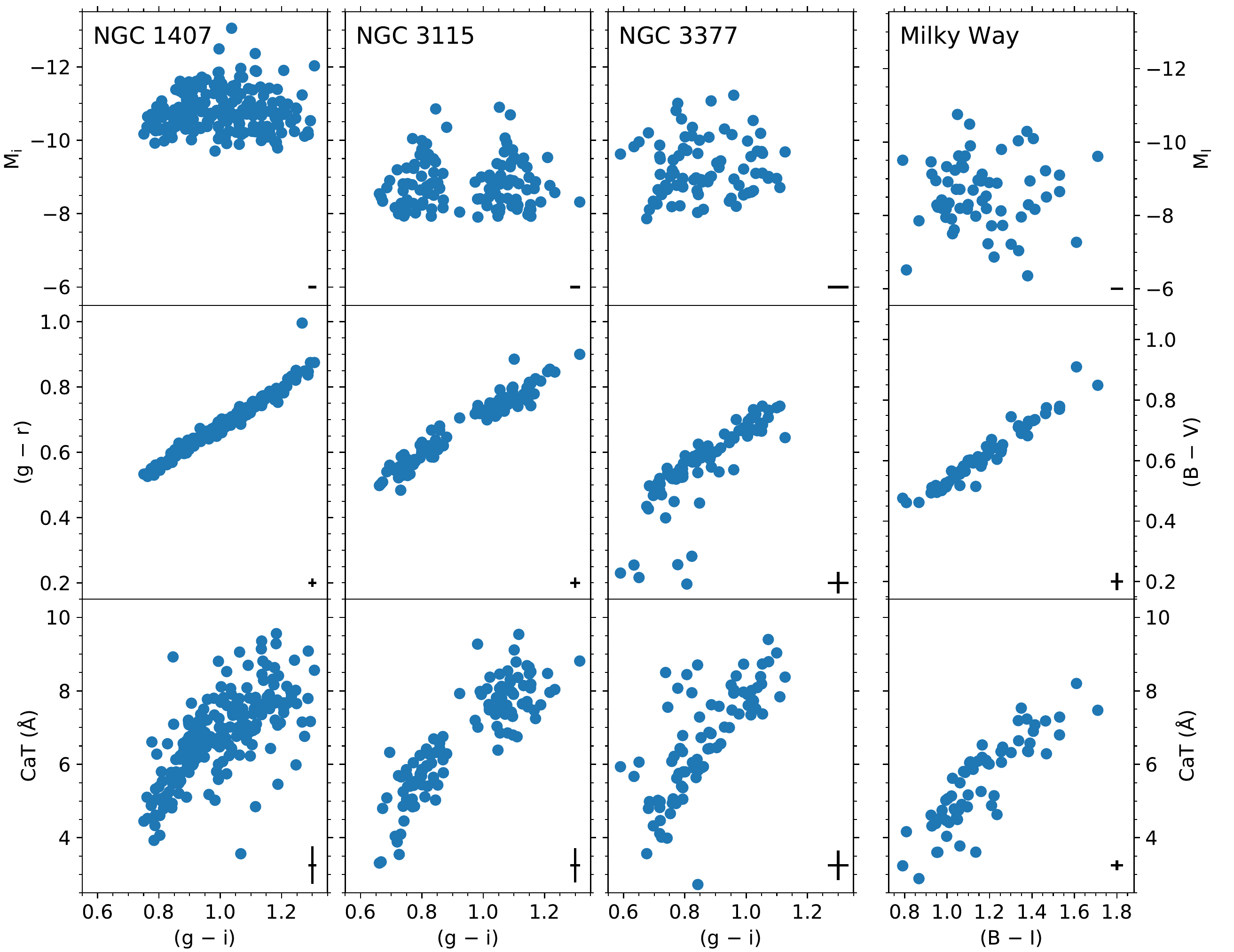}
    \caption{Top row: colour-magnitude diagrams of our GCs in the three SLUGGS galaxies and the MW.
    Middle row: colour-colour plots for the four galaxies.
    Bottom row: colour-CaT strength plots for four galaxies.
    In the lower left of each of the panels the median uncertainties are plotted.
    The ranges of $(B - V)$, $(B - I)$ and $M_{I}$ for the MW correspond to the same range of $(g - r)$,  $(g - i)$ and $M_{i}$ for the other three galaxies.
    There are signifiant differences in the colour and CaT distributions of the four galaxies.
    The GCs in NGC 1407 are significantly more luminous than those in other galaxies.
    The luminosity distributions of the other galaxies are similar to one another although the MW sample extends to fainter GCs.
    \label{fig:colour_mag_CaT}}
\end{figure*}

We apply our MCMC age--metallicity technique to GCs around three galaxies from the SLUGGS survey - NGC 1407, NGC 3115 and NGC 3377.
NGC 1407 is a slowly rotating E0 galaxy with a stellar mass of $4 \times 10^{11}$ M$_{\sun}$, NGC 3115 is a fast rotating S0 galaxy with a stellar mass of $9 \times 10^{10}$ M$_{\sun}$, and NGC 3377 is an E5/6 galaxy with an embedded disc and a stellar mass of $3 \times 10^{10}$ M$_{\sun}$ \citep{2014ApJ...791...80A, 2017MNRAS.464.4611F}.
For the GCs in these three galaxies we use Subaru Suprime-Cam $gri$ photometry presented in \citet{2013MNRAS.428..389P} for NGC 1407 and NGC 3377 and in \citet{2011ApJ...736L..26A} for NGC 3115 and assume an uncertainty in the reddening of $E(B - V) = 0.01$ mag for these GCs.
We use the Keck DEIMOS spectra presented in \citet{2017AJ....153..114F} and use the code described in \citet{2019MNRAS.482.1275U} to measure the CaT strengths before using the empirical relation derived by \citet{2019MNRAS.482.1275U} to convert them into [Fe/H].
For GCs obtained in the SLUGGS survey, the photometric and reddening uncertainties are smaller than for the MW GCs but the uncertainties on the CaT metallicities are larger.
We plot the colours, magnitudes and CaT strengths of GCs in the three SLUGGS galaxies as well as the MW in Figure \ref{fig:colour_mag_CaT} and provide an example posterior distribution for one of the SLUGGS GCs in Figure \ref{fig:NGC1407_S50_corner}.

\begin{figure}
    \includegraphics[width=240pt]{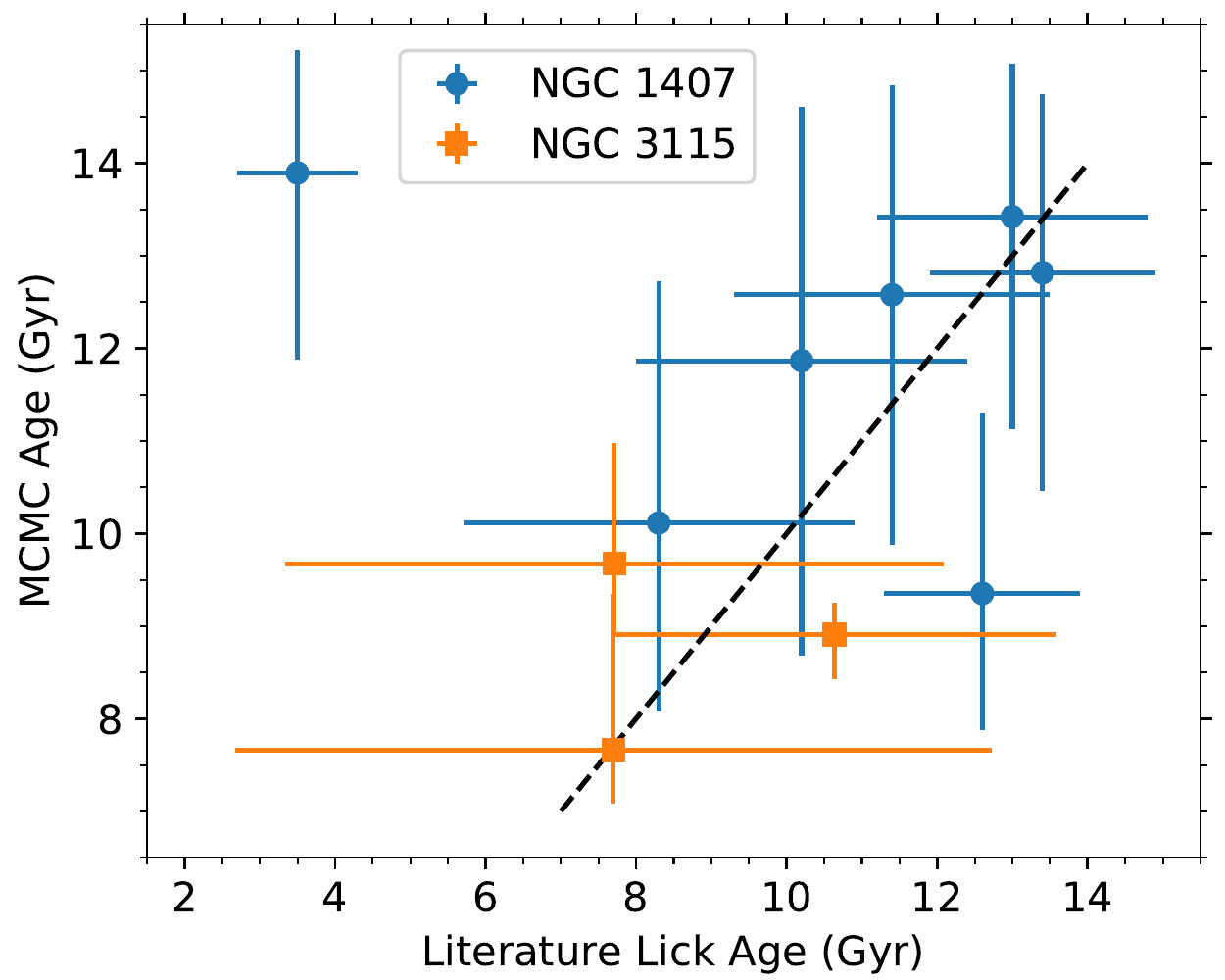}
    \caption{Medians of the MCMC age posteriors versus literature ages for GCs in NGC 1407 (blue circles, from \citealt{2007AJ....134..391C}) and NGC 3115 (orange squares, from \citealt{2002A&A...395..761K} and \citealt{2006MNRAS.367..815N}).
    The dashed line is the one-to-one line.
    There is relatively good agreement for GCs in both galaxies.
    The one NGC 1407 GC with a young literature age was suggested by \citet{2007AJ....134..391C} to be an old GC with a blue HB in agreement with the median of our age posterior.  \label{fig:lit_age_comparison}}
\end{figure}

In Figure \ref{fig:lit_age_comparison} we compare our MCMC age measurements with literature age measurements in NGC 1407 \citep{2007AJ....134..391C} and in NGC 3115 (ages measured by \citealt{2006MNRAS.367..815N} using measurements from \citealt{2002A&A...395..761K}).
Both \citet{2006MNRAS.367..815N} and \citet{2007AJ....134..391C} measured ages by $\chi^{2}$ minimisation of the measured Lick indices \citep{1994ApJS...94..687W} and the stellar population models of \citet{2003MNRAS.339..897T, 2004MNRAS.351L..19T}.
We find good agreement between our age constraints and the literature measurements with the largest disagreement coming from a GC we measure to be old ($13.8_{-2.0}^{+1.3}$ Gyr) but measured to be young by \citet[$3.5 \pm 0.8$ Gyr]{2007AJ....134..391C} but suggested by \citet{2007AJ....134..391C} to be an old GC with a blue HB.

\begin{figure*}
    \includegraphics[width=504pt]{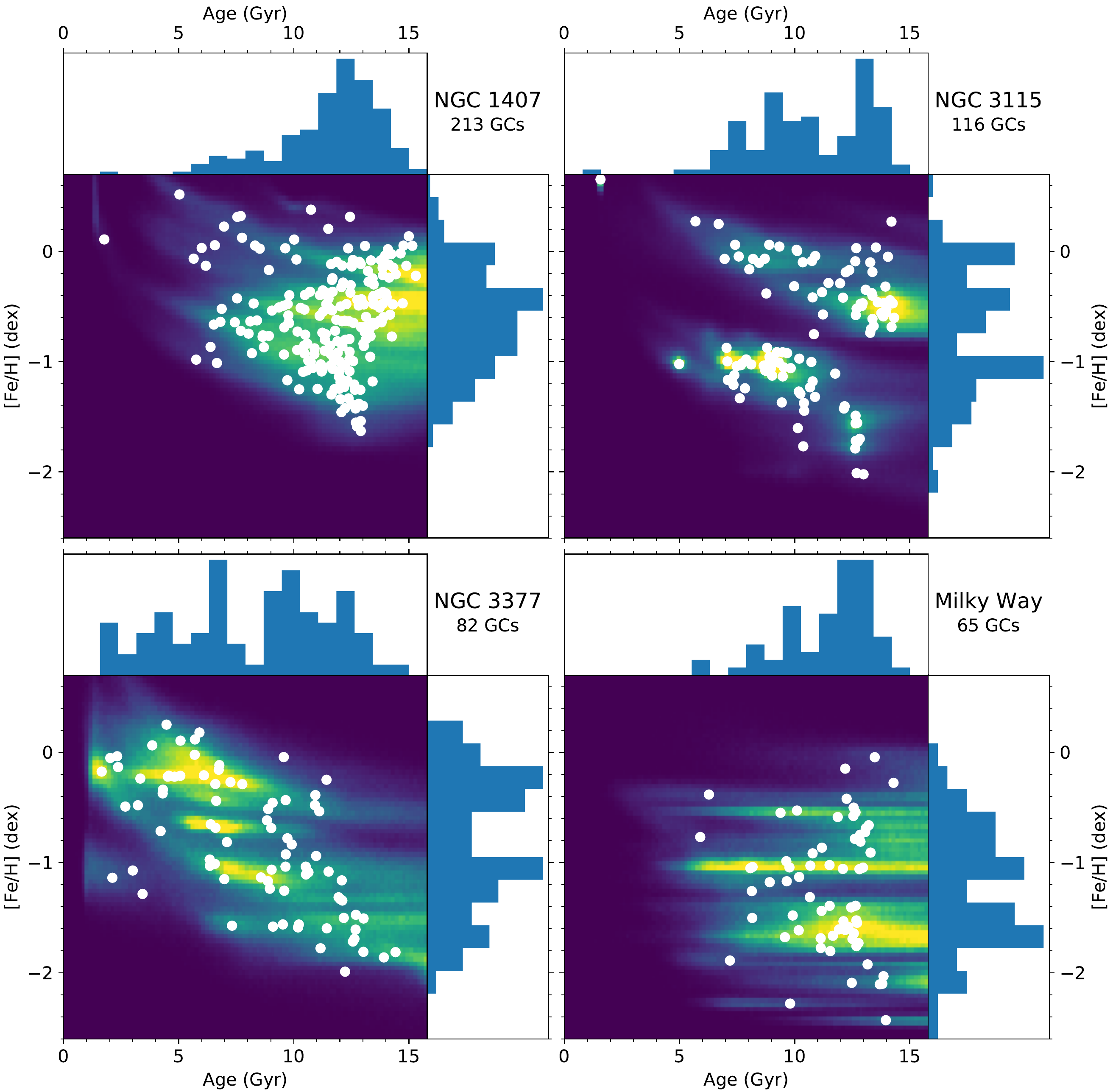}
    \caption{Age--metallicity posterior distribution for GCs around three SLUGGS galaxies and the MW.
    The 2D histograms show the sum of the individual age--metallicity posterior distributions while the white points and the 1D histograms show the medians of the age and metallicity posteriors for each GC.
    The four galaxies show wildly different GC age and metallicity distributions, suggesting different formation histories.
    We note that the photometric and reddening uncertainties are generally larger for the MW GCs than for the SLUGGS GCs while the CaT metallicities are smaller for the MW GCs than for the SLUGGS ones. \label{fig:age_metal_letter}}
\end{figure*}

We plot the age--metallicity posterior distributions for the three galaxies as well as our measurements of MW GCs in Figure \ref{fig:age_metal_letter}.
The four galaxies show wildly different GC age--metallicity distributions.
NGC 1407 shows both an old unimodal age distribution (with a possible tail to intermediate ages) and a unimodal metallicity distribution.
NGC 3115 shows broad age and metallicity distributions with a clear separation between a metal poor age--metallicity sequence and a metal rich sequence.
NGC 3377 also shows broad age and metallicity distributions with a clear age--metallicity sequence.
The MW shows an old unimodal age distribution and a broad metallicity distribution.
For NGC 3115, NGC 3377 and the MW, the mass distributions are broadly similar with median GC masses of $\sim 4 \times 10^{5}$ M$_{\sun}$ and an interquartile range spanning a factor of $\sim 3$.
The more distant NGC 1407 sample, however, has a higher median mass of $\sim 3 \times 10^{6}$ M$_{\sun}$.
Since NGC 1407 displays a strong blue tilt \citep{2006ApJ...636...90H, 2009ApJ...699..254H}, our sample of GCs is deficient in metal poor GCs compared to the galaxy as a whole (see figures 1c and 12 in \citealt{2012MNRAS.426.1475U}).
Red GCs are slightly over represented in our NGC 3377 sample (see figure 1f in \citealt{2012MNRAS.426.1475U}) while the colour distribution of GCs with CaT metallicities closely traces the entire population in NGC 3115 (see figure 1e in \citealt{2012MNRAS.426.1475U}).
The literature metallicity distribution of our sample of MW GCs is consistent with being drawn from the metallicity distribution of the 2010 edition of the \citet{1996AJ....112.1487H, 2010arXiv1012.3224H} catalogue.

By sampling the age posterior distributions we calculate the median ages of each galaxy's GCs which we give in Table \ref{tab:medians} as well as of GCs more or less metal poor than [Fe/H] $= -0.8$ in each galaxy.
[Fe/H] $= -0.8$ is close to the median metallicities of our samples of NGC 3115 and NGC 3377 as well as neatly dividing the two age--metallicity sequences in NGC 3115.
If we use the median metallicities for each galaxy's GCs, we get identical median ages for the two subpopulations as for the [Fe/H] $= -0.8$ split.
The median ages of the different galaxy populations and metallicity subpopulations reinforce the diversity of age distributions.
NGC 3377 has the youngest median age while NGC 1407 and the MW have the oldest.
In both NGC 1407 and the MW the ages of the two metallicity subpopulations are consistent, in contrast to NGC 3115, where the metal poor GCs are slightly younger than the metal rich ones, and to NGC 3377, where the metal rich GCs are significantly younger than the metal poor ones.
At high metallicity NGC 1407's and the MW's GCs are mostly old, NGC 3115's span a wide range of ages and NGC 3377 metal rich GCs appear mostly young.
At lower metallicities the differences between galaxies are less obvious but unlike the other galaxies NGC 3377 shows evidence for a handful of $\sim 2$ Gyr GCs at [Fe/H] $\sim -1$.
Despite NGC 3115 and NGC 3377 showing consistent metallicity distributions (a Kolmogorov-Smirnov test returns a p-value of 0.55), the two galaxies show strongly inconsistent age distributions (a KS test p-value of $1 \times 10^{-6}$).
This highlights the importance of age in studying the formation history of these GCs systems; an analysis based solely of metallicity would conclude the two GC systems were similar.
On the other hand, the MW and NGC 1407 show similar age distributions (a KS test p-value of 0.46) but the NGC 1407 GCs are on average much more metal rich, even after accounting for the effects of NGC 1407's blue tilt.

\begin{table}
    \caption{GC age medians}
    \label{tab:medians}
    \begin{tabular}{ccccccc}
        Galaxy & \multicolumn{2}{c}{Full} & \multicolumn{2}{c}{[Fe/H] $< -0.8$} & \multicolumn{2}{c}{[Fe/H] $> -0.8$}  \\
        & [Gyr] & $N$ & [Gyr] & $N$ & [Gyr] & $N$ \\ \hline
        NGC 1407 & $11.9_{-0.3}^{+0.3}$ & 213 & $11.7_{-0.4}^{+0.4}$ & 74 & $12.0_{-0.3}^{+0.3}$ & 139 \\
        NGC 3115 & $10.9_{-0.4}^{+0.4}$ & 116 & $9.7_{-0.4}^{+0.5}$ & 56 & $12.3_{-0.5}^{+0.5}$ & 60 \\
        NGC 3377 & $8.0_{-0.4}^{+0.5}$ & 82 & $10.3_{-0.7}^{+0.7}$ & 42 & $6.1_{-0.5}^{+0.5}$ & 40 \\
        Milky Way & $11.7_{-0.6}^{+0.5}$ & 65 & $11.6_{-0.6}^{+0.6}$ & 48 & $12.1_{-1.1}^{+1.0}$ & 17 \\ \hline
    \end{tabular}
\end{table}

\section{Discussion} \label{sec:discussion}

The good agreement between our Local Group GC ages and literature ages (see Figure \ref{fig:age_comparison}) indicates that the effects of HB morphology or non-solar abundance ratios on our ages are likely relatively small.
In the MW and other galaxies, there is a strong correlation between the morphology of the HB and GC mass at fixed metallicity, with more massive GCs having bluer HBs \citep[e.g.][]{2006A&A...452..875R, 2010A&A...517A..81G, 2017MNRAS.464..713P, 2018MNRAS.481.3313P}.
This correlation is likely due to a correlation between GC mass and the internal spread of He abundance, with more massive GCs having larger ranges of He abundance \citep[e.g.][]{2002A&A...395...69D, 2018MNRAS.481.5098M}.
Thus, the relative ages of GCs of similar masses should be reliable, suggesting that the differences between the NGC 3115, NGC 3377 and MW age distributions are real.
If our ages were significantly affected by the morphology of the HB, we would expect the more massive GCs in NGC 1407 to appear younger yet their ages are consistent with those in the MW.
We plan to quantify the effects of HB morphology and different abundance patterns on our ages in future work.
We note that our ages are likely more reliable in a relative sense than absolute, especially at fixed metallicity.
Our GC age and metallicity measurements are also in agreement with previous spectroscopic studies (see Figure \ref{fig:lit_age_comparison}) in the blue of NGC 3115 \citep{2002A&A...395..761K, 2006MNRAS.367..815N} and NGC 1407 \citep{2007AJ....134..391C} and qualitatively with the optical-near-infrared imaging study of NGC 3377 by \citet{2011A&A...525A..20C}.
We note that our samples are $\sim 10 \times$ larger than the previous spectroscopic studies of these galaxies and achieved with significantly shorter exposure times ($\sim 2$ h per mask for our observations compared to 3.5 h for \citealt{2002A&A...395..761K} and 7 h for \citet{2007AJ....134..391C}).

The diversity in GC ages and metallicities both within and between galaxies agrees with the range of colour--CaT relationships observed by \citet{2015MNRAS.446..369U}, the variation in colour-metallicity relationship seen by \citet{2019ApJ...879...45V} and by the variation in GC colour-colour relationship with environment seen by \citet{2016ApJ...829L...5P}.
Further evidence for this diversity in stellar populations is seen with the various age distributions observed in previous spectroscopic studies which find uniformly old ages in some galaxies (e.g. NGC 1399, \citealt{2001ApJ...563L.143F}, NGC 4365, \citealt{2005AJ....129.2643B, 2012MNRAS.427.2349C}) but not others (e.g. NGC 1316, \citealt{2001MNRAS.322..643G, 2018MNRAS.479..478S}, NGC 5128, \citealt{2004ApJ...602..705P, 2008MNRAS.386.1443B}).
This diversity implies that a single colour--metallicity relation is not a good description of most GC systems as has been assumed by most previous studies.
In particular, the assumption that different mass galaxies follow the same colour-metallicity relationship is incorrect since lower mass galaxies tend to host younger GCs \citep[e.g.][this work]{2006MNRAS.372.1259S, 2006ApJ...646L.107C, 2008A&A...489.1065M, 2014AJ....147...71P}.

\subsection{Galaxy assembly histories}
We can qualitatively interpret the assembly histories of the 3 SLUGGS galaxies using the framework of \citet{2019MNRAS.486.3134K} in an analogous manner to the study of the MW by \citet{2019MNRAS.486.3180K} using the E-MOSAICS simulations \citep{2018MNRAS.475.4309P, 2019MNRAS.486.3134K} \footnote{The analysis of \citet{2019MNRAS.486.3134K, 2019MNRAS.486.3180K} was tailored to central MW mass galaxies but the general principles should be applicable to any galaxy massive enough to host a populous GC system.}.
As described by \citet{2019MNRAS.486.3134K}, the evolution of the stellar age--metallicity relation of galaxies (as predicted by cosmological simulations such as EAGLE, \citealt{2015MNRAS.446..521S}, \citealt{2015MNRAS.450.1937C}) can be used to estimate what mass of galaxy a GC of a given age and metallicity likely formed in.
By using the evolution of galaxies in age--metallicity space, it is possible to connect GCs to a likely progenitor.
The GCs associated with the most massive progenitor can be linked to the in situ population while the GCs associated with lower mass progenitors were likely accreted.
The youngest GC associated with a particular lower mass progenitor places an upper limit on when that progenitor was accreted as cluster formation halts in the lower mass progenitor when it is accreted.
By using the observed scaling between the number of GCs and host galaxy halo mass \citep[e.g.][]{1997AJ....114..482B, 2009MNRAS.392L...1S, 2017ApJ...836...67H} it is possible to estimate the number of GCs formed by a galaxy with a given age and metallicity.
By comparing this expected number with the observed number of GCs, the number of progenitors of a given mass can be estimated.
As noted by \citet{2019MNRAS.486.3134K}, this analysis is most sensitive to unequal mass mergers between redshifts of $z = 1$ and 2.5 where the accreted galaxies are massive enough to form GCs (stellar masses of $\sim 10^{8}$ M$_{\sun}$ and above).
Above a redshift of $z = 2.5$, the steep age--metallicity relations of different masses galaxies become difficult to distinguish from one another especially with realistic age uncertainties, while below $z = 1$ cosmic GCs formation declines rapidly \citep{2019MNRAS.486.5838R}.

\subsubsection{NGC 1407}
That most GCs of NGC 1407 are old (ages $> 10$ Gyr) implies that NGC 1407 formed most of its stellar mass early.
The likely tail to younger ages in NGC 1407 suggests that NGC 1407 assembled over an extended period of time while the wide range of metallicities of these younger GCs suggests NGC 1407 was built from a wide range of mass progenitors.
We note that the age uncertainties of our NGC 1407 GCs are smaller than in the MW due to the smaller uncertainties on the NGC 1407 photometry and reddening.  
Stellar population studies of NGC 1407's field star population \citep[e.g.][]{2008MNRAS.385..675S, 2018MNRAS.480.3215J, 2019ApJ...878..129F} paint a consistent picture with old ages (11 to 13 Gyr) and high [$\alpha$/Fe] values (suggestive of rapid star formation).
The kinematics \citep[e.g.][]{2014ApJ...791...80A, 2018MNRAS.480.3215J, 2010ApJ...725.2312O, 2019ApJ...878..129F} and the strong field star metallicity gradient \citep{2008MNRAS.385..675S, 2014MNRAS.442.1003P, 2019ApJ...878..129F} are suggestive of a large accretion fraction for NGC 1407's stellar mass.
This is in line with cosmological simulations of galaxy formation \citep[e.g.][]{2010ApJ...725.2312O, 2016MNRAS.458.2371R, 2017MNRAS.464.1659Q, 2018MNRAS.479.4760F} which predict the most massive galaxies, such as the brightest group galaxy NGC 1407, form most of their stellar mass early but assemble their mass late.

\subsubsection{NGC 3115}
The two age--metallicity sequences of NGC 3115 can be identified with a more metal rich population that formed in situ and a more metal poor population that formed either in lower mass galaxies that merged with NGC 3115 or from gas that was brought in by lower mass galaxies accreting onto NGC 3115.
The two sequences span similar age ranges although the in situ GCs likely formed earlier on average than the accreted GCs.
The large, clear gap in metallicity between the two sequences suggests a substantial difference in stellar mass between the progenitors.
The roughly equal numbers of metal poor and metal rich GCs suggest equal contributions from in situ and accreted GCs under the assumption that the low metallicity branch is dominated by accreted GCs although some fraction of the old metal poor GCs likely formed in situ.
Since the number of GCs scales with galaxy mass, this suggests a number of lower mass galaxies were accreted onto NGC 3115 to build up its GC system.
The ages of the in situ GCs show agreement with studies of NGC 3115's field star populations \citep{2006MNRAS.367..815N, 2016A&A...591A.143G, 2019MNRAS.487.3776P} with an old (10 to 13 Gyr), $\alpha$-enhanced spheroid and a disc with a star forming history declining from a peak at old ages to 5 Gyr and younger.
The bimodal nature of NGC 3115's formation is supported by the bimodal halo star metallicity distribution measured by \citet{2015ApJ...800...13P}, with the peaks of the halo field star distribution coinciding with those of the GCs.
The in situ build-up of NGC 3115 bulge and disc without late major mergers and the build up of NGC 3115's halo by many minor mergers is supported by the joint stellar population-dynamical study of \citet{2019MNRAS.487.3776P}.

\subsubsection{NGC 3377}
The large age spread in NGC 3377 GCs suggests an extended formation history while the lack of old, metal rich GCs compared to the other SLUGGS galaxies and the MW suggests a late peak in its star formation history relative to the other galaxies in this study.
However, the peak of GC formation is close to the expected peak of star formation for a galaxy of its mass \citep[e.g.][]{2017MNRAS.464.1659Q}.
The wide range of metallicities at intermediate ages shows the contribution of accreted galaxies with the trio of GCs with ages of $\sim 4$ Gyr and metallicities of [Fe/H] $\sim -1.2$ suggests the relatively recent accretion of a dwarf galaxy. 
In agreement with wide range GC ages in NGC 3377, \citet{2015MNRAS.448.3484M} measured a luminosity weighted age of $7 \pm 1$ Gyr and a mass weighted age of $11 \pm 1$ Gyr for the field star population of NGC 3377.

\subsubsection{Milky Way}
The assembly history of the MW has already been studied using its GC ages and metallicities by \citet{2019MNRAS.486.3180K} and we do not attempt to repeat their analysis here.
We do note that within our admittedly large uncertainties, our age and metallicity measurements agree with literature age-metallicity relationships for the MW \citep[e.g.][]{2010MNRAS.404.1203F, 2011ApJ...738...74D, 2013MNRAS.436..122L, 2019MNRAS.486.3180K}.
\citet{2019MNRAS.486.3180K} concluded that the MW assembled relatively early for a galaxy of its mass, in line with the evidence from the MW's field stars \citep[e.g.][]{2014ApJ...781L..31S, 2018MNRAS.477.5072M}.
The prediction by \citet{2019MNRAS.486.3180K} that the MW's accretion history has been dominated by three relatively massive galaxies has been supported with the association of a large number of the MW GCs with a similar number of progenitors identified in \textit{Gaia} and earlier data \citep[e.g.][]{2017A&A...598L...9M, 2018ApJ...862...52S, 2018ApJ...863L..28M, 2018Natur.563...85H, 2019MNRAS.488.1235M, 2019arXiv190608271M}.
The ages and metallicities of the MW's halo and thick disc stars \citep[e.g.][]{2012A&A...538A..21S, 2019NatAs.tmp..407G} show good agreement with the literature values for MW GCs.
The large population of young and intermediate age GCs in the Magellanic Clouds is also consistent with the star formation history in these galaxies \citep[e.g.][]{2009AJ....138.1243H, 2008AJ....135..836C, 2014AJ....147...71P, 2015MNRAS.448.1863B}.

Taken together the assembly histories implied by the age--metallcity distributions of the four galaxies agree with the evidence from other observables.
The median ages of the metal rich GCs in the three SLUGGS galaxies ($12.0 \pm 0.3$, $12.3 \pm 0.5$ and $6.1 \pm 0.5$ Gyr for NGC 1407, NGC 3115 and NGC 3377 respectively) agree well with the ages of the majority of field stars (11 to 13, 10 to 13 and 7 to 11 Gyr respectively), implying that the metal rich GCs in these galaxies formed on average at the same time as most of the field stars.
Since the ratio of mass in GCs to field star mass at fixed metallicity strongly declines with increasing metallicity \citep[e.g.][]{2002AJ....123.3108H, 2012A&A...544L..14L, 2017A&A...606A..85L}, we expect the ages of the metal rich GCs to more closely trace the field star age distribution than the ages of all GCs.
This agreement supports models of GC formation \cite[e.g.][]{2015MNRAS.454.1658K, 2017ApJ...834...69L, 2018MNRAS.475.4309P, 2019arXiv190611261M} where GC formation is the natural outcome of intense star formation and galaxies with high star formation rates at later times can form younger GCs \citep{2015MNRAS.454.1658K}.
Since the GC formation is biased to the most intense period of star formation and the population of GCs we observe today is the product of both GC formation and destruction, while we expect a connection between the field star age distribution and the GC age distribution, we do not expect that the GC age distribution to be exactly the same as the field star age distribution in an analogous manner to the relationship between the GC metallicity distribution and the field star metallicity distribution.
The range of ages we observe in both metal poor and metal rich GCs indicates that GC formation is a process that occurs across cosmic time and rules out models that predict that GC formation requires special conditions in the early Universe.
The ongoing formation of GCs is observed today with the formation of young, massive star clusters in galaxies with extreme star formation rates \citep[e.g.][]{1992AJ....103..691H, 1993AJ....106.1354W, 2004A&A...416..467M, 2006A&A...448..881B, 2010ARA&A..48..431P}.

That the lowest mass galaxy of the three SLUGGS galaxies (NGC 3377) has the youngest GCs on average while the most massive (NGC 1407) has the oldest GCs  is in line with galactic downsizing \citep[e.g.][]{1992MNRAS.254..601B, 1996AJ....112..839C, 2005ApJ...621..673T, 2009MNRAS.397.1776F} although there is a diversity of galaxy assembly histories at fixed stellar mass.
This diversity at fixed stellar masses is illustrated by the fact that the MW hosts an older GC system on average than NGC 3115 or NGC 3377 despite the its stellar mass ($6 \times 10^{10}$ M$_{\sun}$ e.g. \citealt{2016ARA&A..54..529B}) is between the SLUGGS galaxies ($9 \times 10^{10}$ and $3 \times 10^{10}$ M$_{\sun}$ respectively).
Understanding how much scatter exists in the age distributions of GC systems and how much variation in galaxy assembly histories exists, both with stellar mass and at fixed mass, requires a larger sample of galaxies.

\section{Conclusions} \label{sec:conclusions}

In this work we have used a novel technique to combine age-independent metallicities from the CaT with optical photometry to measure the ages of GCs around the MW and three galaxies from the SLUGGS survey.
We find good agreement between our ages and those from literature (Figures \ref{fig:age_comparison} and \ref{fig:lit_age_comparison}).
We find a diversity of age--metallicity distributions, implying different galaxy assembly histories (Figure \ref{fig:age_metal_letter}).
We note that both ages and metallicities are required to distinguish between different formation histories as galaxies with similar metallicity distributions can have wildly different age distributions (i.e. NGC 3115 and NGC 3377).
The assumption that all GCs in all galaxies follow the same colour-metallicity relationship, long used in extragalactic GC studies, needs to be discarded.
Our observed connection between the median metal rich GC age and field star age shows GC formation traces the peaks of star formation.
We observe a wide range of GCs ages, demonstrating that GC formation is a process ongoing to the present day.
Taken together, these findings support models where GC formation is a natural outcome of star formation and reject models where GC formation requires special conditions in the early universe.
We used the age--metallicity distributions of the three SLUGGS galaxies to study their formation and assembly histories.
We find that NGC 1407 formed its stellar mass early but had an extended assembly history, that NGC 3115's assembly was dominated by a number of unequal mass mergers and that NGC 3377 formed its stellar mass relatively late, as expected for a galaxy of its stellar mass.
In future work, we will extend this age--metallicity analysis to the remaining SLUGGS galaxies.
We will quantitatively compare the observed age--metallicity distributions with the simulated GC systems of galaxies with the same stellar masses from E-MOSAICS and from other cosmological models of GC formation in order to constrain the assembly histories of the SLUGGS galaxies.

\section*{Acknowledgements}
We thank the anonymous referee for their helpful suggestions which improved the manuscript.
We thank Alexa Villaume, Ana Chies Santos, Ricardo Schiavon, Marta Reina-Campos and Diederik Kruijssen for useful discussions.
CU, JP and NB gratefully acknowledge financial support from the European Research Council (ERC-CoG-646928, Multi-Pop).
JPB gratefully acknowledges support from the National Science Foundation grants AST-1518294 and AST-1616598.
DF gratefully thanks the Australian Research Council for financial support via DP160101608.
AJR was supported by National Science Foundation grants AST-1515084 and AST-1616710, and as a Research Corporation for Science Advancement Cottrell Scholar.
JS acknowledges support from NSF grant AST-1514763 and a Packard Fellowship.
NB gratefully acknowledges financial support from the Royal Society (University Research Fellowship).
The authors wish to recognize and acknowledge the very significant cultural role and reverence that the summit of Maunakea has always had within the indigenous Hawaiian community.
We are most fortunate to have the opportunity to conduct observations from this mountain.

This work made use of \textsc{numpy} \citep{numpy}, \textsc{scipy} \citep{scipy}, \textsc{matplotlib} \citep{matplotlib} and \textsc{corner} \citep{corner} as well as \textsc{astropy}, a community-developed core Python package for astronomy \citep{2013A&A...558A..33A} and \textsc{Python-FSPS} \citep{dan_foreman_mackey_2014_12157}.

\bibliographystyle{mnras}
\bibliography{bib}{}

\appendix
\section{Example posterior distributions}

\begin{figure*}
    \includegraphics[width=504pt]{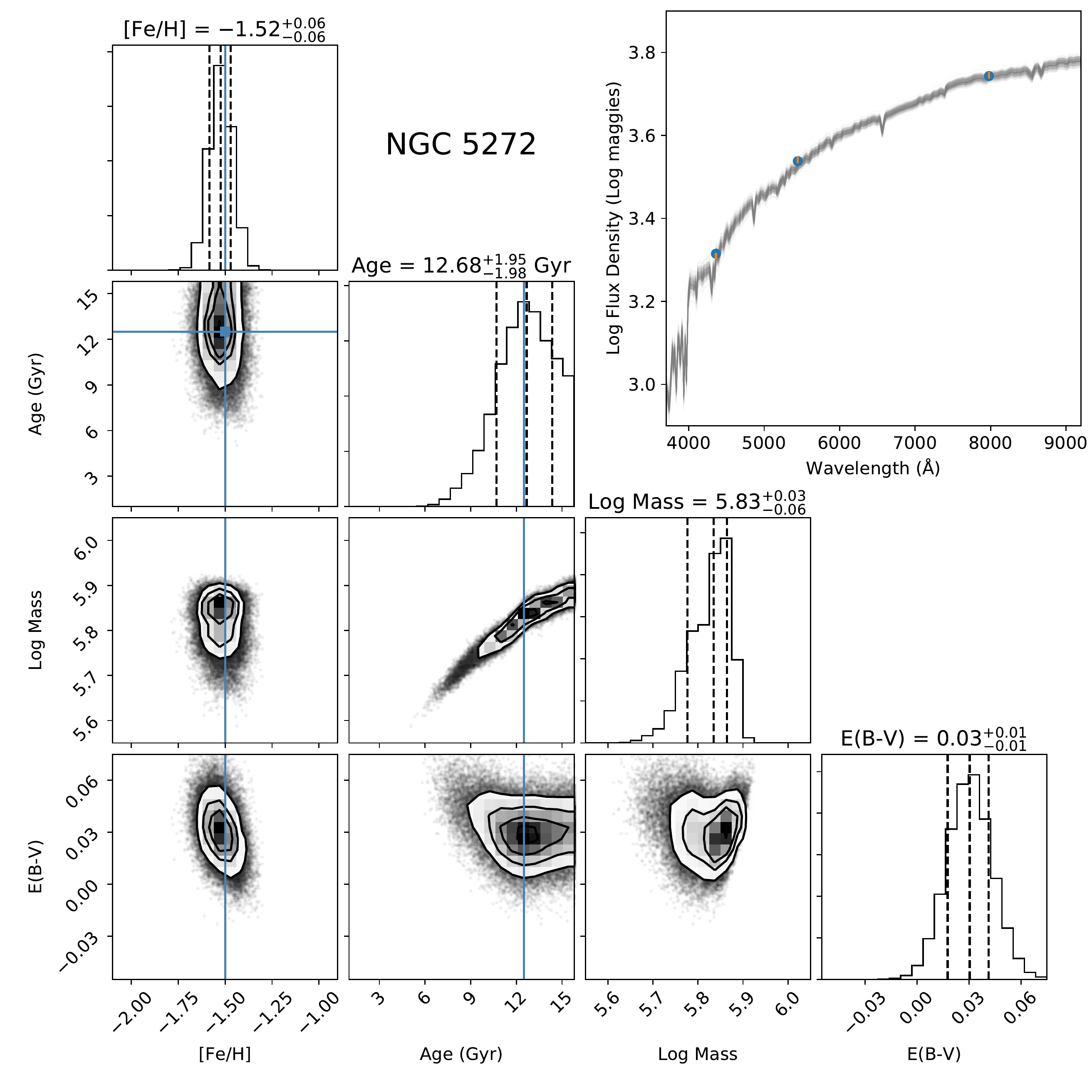}
    \caption{Posterior distributions of metallicity, age, mass and reddenings for the MW GC NGC 5272.
    The metallicity from the 2010 edition of the \citet{1996AJ....112.1487H, 2010arXiv1012.3224H} catalogue and resolved colour-magnitude age from \citet{2010ApJ...708..698D} are over plotted in blue.
    In the top left we plot the spectral energy distributions calculated by \textsc{fsps} for 256 points drawn at random from the posterior distribution in grey, the median colours of the posterior distribution as blue circles and the observed photometry as orange error bars.
    \label{fig:NGC5272_corner}}
\end{figure*}

\begin{figure*}
    \includegraphics[width=504pt]{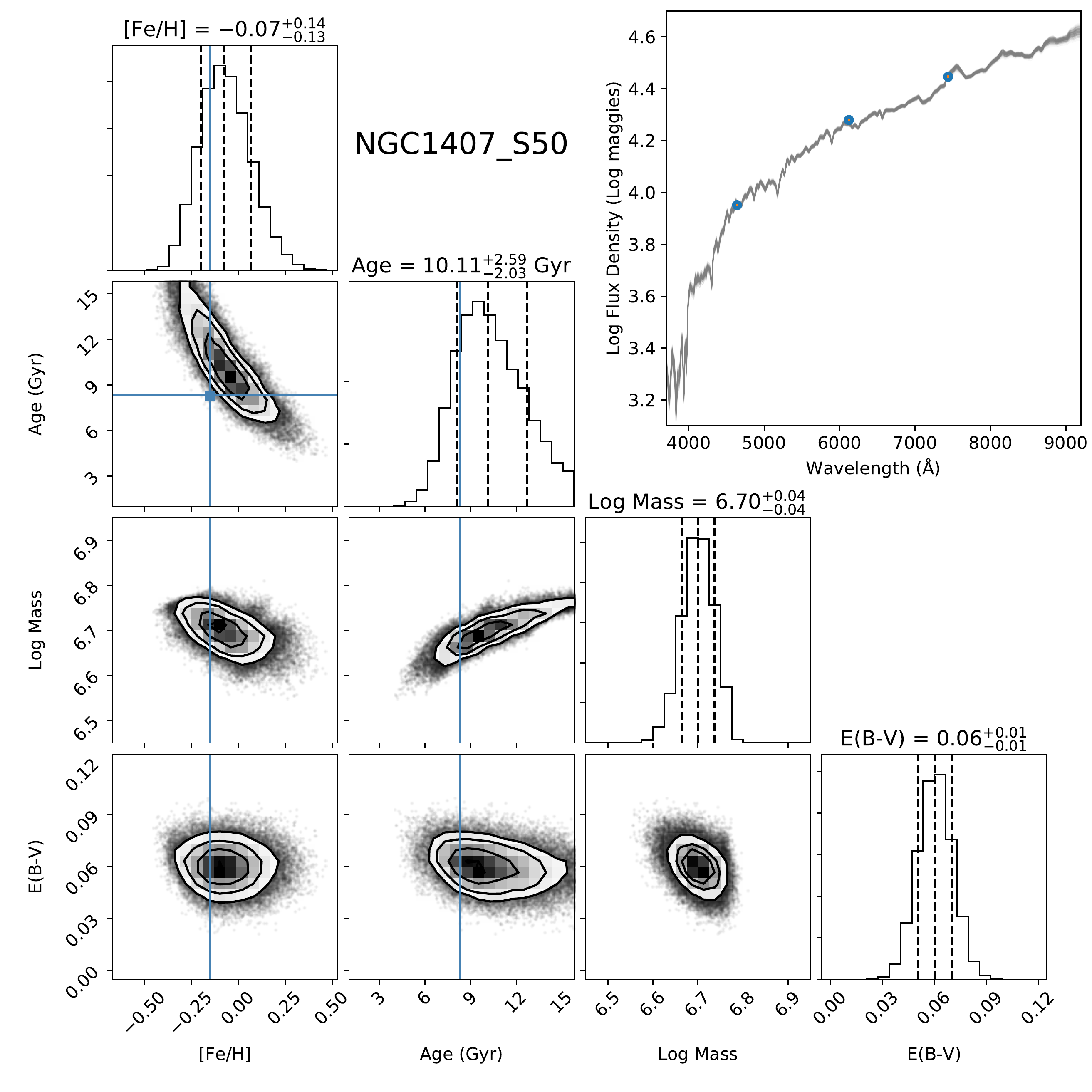}
    \caption{Posterior distributions of metallicity, age, mass and reddenings for the NGC 1407 GC NGC1407\_S50.
    The metallicity and age based on the analysis of Lick indices by \citet{2007AJ....134..391C} are over plotted in blue.
    In the top left we plot the spectral energy distributions calculated by \textsc{fsps} for 256 points drawn at random from the posterior distribution in grey, the median colours of the posterior distribution as blue circles and the observed photometry as orange error bars.
    \label{fig:NGC1407_S50_corner}}
\end{figure*}

\end{document}